\documentclass[aps,prl,fleqn,showpacs,showkeys,twocolumn,nofootinbib,
preprintnumbers,superscriptaddress]{revtex4-1}
\usepackage{amsmath,amsfonts,amssymb,amscd,amsxtra,amsthm}
\usepackage{graphicx}  
\usepackage{epstopdf}
\usepackage{color}
\usepackage{dcolumn}  
\usepackage{bm}          
\usepackage{multirow}
\usepackage{slashed}
\usepackage[utf8]{inputenc}

\usepackage[normalem]{ulem} 

\begin{document}

\preprint{INHA-NTG-01/2016}
\title{Pion mean fields and heavy baryons}

\author{Ghil-Seok Yang}
\affiliation{Department of Physics, Soongsil University, Seoul 06978,
  Republic of Korea}
\email{ghsyang@ssu.ac.kr}

\author{Hyun-Chul Kim}
\affiliation{Department of Physics, Inha University, Incheon 22212,
  Republic of Korea}
\email{hchkim@inha.ac.kr}

\affiliation{School of Physics, Korea Institute for Advanced Study
  (KIAS), Seoul 02455, Republic of Korea}

\author{Maxim V. Polyakov}
\affiliation{Institut f\"ur Theoretische Physik II, Ruhr-Universit\"at
  Bochum, D--44780, Bochum, Germany} 
\email{maxim.polyakov@tp2.ruhr-uni-bochum.de}

\affiliation{Petersburg Nuclear Physics Institute, Gatchina,
  St. Petersburg 188 350, Russia}

\author{Micha{\l{}} Prasza{\l{}}owicz}
\affiliation{M. Smoluchowski Institute of Physics, Jagiellonian
  University, {\L}ojasiewicza 11, 30-348 Krak{\'o}w, Poland}
\email{michal.praszalowicz@uj.edu.pl}

\begin{abstract} 
We show that the masses of the lowest-lying heavy baryons can be very
well described in a pion mean-field approach. We consider a heavy
baryon as a system consisting of the $N_c-1$ light quarks that induce
the pion mean field, and a heavy quark as a static color source under
the influence of this  mean field. In this approach we derive a number
of \textit{model-independent} relations and calculate the heavy baryon 
masses using those of the lowest-lying light baryons as input. The
results are in remarkable agreement with the experimental data. In
addition, the mass of the $\Omega_b^*$ baryon is predicted. 
\end{abstract}



\maketitle

\textit{Heavy baryons and pion mean-field}.-- In a naive
quark model a heavy baryon consists of a heavy quark and two light
quarks. When the mass of the heavy quark  $m_Q\to \infty$,  the spin
of the heavy quark $\bm  S_Q$ is conserved, which indicates that the
spin of the light-quark degrees of freedom is also conserved:
$\bm{S}_{\mathrm{L}} \equiv  \bm{S}-\bm{S}_Q$~\cite{Isgur:1989vq,
Isgur:1991wq, Georgi:1990um}. Because of this heavy-quark spin
symmetry, the total spin of the light quarks can be considered as a
good quantum number. This suggests that in the first
approximation a heavy baryon can be viewed as a bound state of a heavy
quark and a diquark. Thus, the flavor $SU(3)_{\mathrm{f}}$
representations of the lowest-lying heavy baryons are: $\bm{3}\otimes 
\bm{3}=\overline{\bm{3}} \oplus \bm{6}$, of which the anti-triplet
has $S_{\mathrm{L}}=0$ and total $S=1/2$ and the sextet  has
$S_{\mathrm{L}}=1$ with $S=1/2$ and $S=3/2$. Since in the limit  
$m_Q\to \infty$ the heavy quark inside a
heavy baryon can be regarded as a static color source, the dynamics
of heavy baryons is governed by the light quarks.  

It is clear that the complete description of heavy baryons
requires more involved treatment of light quarks.
In the present Letter we propose to describe the dynamics
of the light subsystem in a heavy baryon within a mean-field approach
with a \textit{hedgehog}~\cite{Skyrme:1961vq} symmetry, motivated by   
Ref.~\cite{Diakonov:2010tf}. Mean field approximations provide
often a simple physical picture, so that they have been widely applied
in a variety of fields in physics: Thomas-Fermi approximation in atomic
physics, Ginzburg-Landau theory for superconductivity, Bethe method in
statistical physics, shell models in nuclear physics, to name a few.
In the seminal papers~\cite{Witten:1979kh} 
Witten has argued that,
to the leading order in $1/N_c$ expansion, the 
lowest-lying light baryons can be also viewed as  bound states of $N_c$
\textit{valence} quarks in a mean field. 
In the limit of the large number of colors ($N_c$), the lowest-lying
light baryons consist of $N_c$ \textit{valence} quarks that produce an 
effective pion mean field, which arises from the vacuum
polarization. The $N_c$ valence quarks are  influenced by this pion
mean field. The chiral-quark soliton model ($\chi$QSM) is constructed, 
based on this
picture~\cite{Diakonov:1987ty,Christov:1995vm,Diakonov:1997sj}.       
This mean field and a \textit{hedgehog} symmetry allow one to derive
the effective collective Hamiltonian that includes an explicit
breaking of $\mathrm{SU(3)_{\mathrm{f}}}$ symmetry. The Hamiltonian
involves the dynamical coefficients, which can be computed explicitly  
within the $\chi$QSM~\cite{Blotz:1992pw} in terms of the relativistic
single particle quark states in the soliton background
configuration. What will be  important in the following is that the
quark-soliton configuration has a trivial color structure: 
it consists of $N_c$ copies of a colorless soliton. This means that
in the leading order all dynamical coefficients -- so called
\textit{moments of inertia} -- of the effective Hamiltonian are
proportional to $N_c$.  If the mean field is generated
- as in the present case - by $N_c-1$, rather than by $N_c$ quarks,
these coefficients have to be appropriately rescaled.

In the large $N_c$ limit, heavy baryons consist of a heavy quark
and $N_c-1$ light quarks rather than a diquark. In this limit
the $N_c-1$ valence quarks produce again the pion
mean field and the system can be described as a quark-soliton.
In the case of the light baryons the SU(3)$_{\mathrm{f}}$ space
of the effective Hamiltonian is a subject to a constraint imposed by
the $N_c$ \textit{valence} quarks: $Y'=N_c/3$ that selects the lowest
allowed representations: {\bf 8} and {\bf 10}. In the heavy baryon
case the constraint is modified $Y'=(N_c-1)/3$ due to the presence of
the $N_c-1$ \textit{valence} quarks, and the lowest allowed
representations are $\overline{\bm{3}}$ and $\bm{6}$. The model 
predicts the structure of the symmetry breaking and allows one to 
compute numerical values of the dynamical coefficients.  

In the general framework of this mean-field picture, one can extract
the moments of inertia from the experimental data on the masses of the
lowest-lying light-quark baryons without 
relying on any model calculation~\cite{Adkins:1984cf,
  Diakonov:1997mm}. Such an analysis has been performed recently 
in Ref.~\cite{Yang:2010fm} and the dynamical parameters have been
determined with high accuracy.  In the present work we shall use these
values for the description of heavy baryon masses.  Additionally, we
shall introduce a spin-spin interaction~\cite{Zeldovich} to remove
spin $1/2$ and $3/2$ degeneracy of the sextet states. The hyperfine
coupling -- the only parameter undetermined from the light sector --
will be fixed from the experimental data.

\vspace{0.2cm}

\textit{Collective Hamiltonian}.-- The SU(3) soliton is constructed in
terms of a \textit{hedgehog} \cite{Skyrme:1961vq} Ansatz, 
which couples three first Gell-Mann matrices with a unit space
vector $\vec{n} \cdot  \vec{\lambda}$.
It is an extended object and therefore its quantization is similar
the textbook quantization of a rigid body. This requires to
identify the zero modes, which in the  case of a \textit{hedgehog} correspond
to the space rotations and the rotations in the flavor space.
Since the soliton lives in the isospin SU(2) subspace of the SU(3) 
group, the rotation along the hypercharge axis is not dynamical
and the corresponding generalized momentum produces a constraint:
the only representations of the SU(3) group that are allowed must
contain states with hypercharge $Y'$ (called \textit{right} hypercharge)
whose actual value depends on
the number of valence quarks. Moreover, the isospin of the states with
hypercharge equal to $Y'$ is equal to the soliton spin. Details can be found in
Refs.~\cite{Blotz:1992pw,SU(3)}.
A general form of the collective rotational
Hamiltonian for the light-quark takes therefore a form of the
quantized symmetric top rotating in the flavor SU(3) space:
\begin{align}
H_{\left(p,\, q\right)}^{\mathrm{rot}} 
 =  M_{\mathrm{sol}}+
\frac{1}{2I_{1}}\sum_{i=1}^{3}\hat{J}_{i}^{2} +
\frac{1}{2I_{2}}\sum_{a=4}^{7}\hat{J}_{a}^{2},
\label{eq:Hcoll}
\end{align}
where $\hat{J}_{i}$ are generators of the SU(3) group,
whose first three components correspond to the soliton spin. 
$I_{1,2}$ stand for the moments of inertia and $M_{\rm sol}$ is
a classical soliton mass. 
Note that $\hat{J}_{8}$ corresponding to $Y^{\prime}$
 does not appear in Eq.~(\ref{eq:Hcoll}).
It is, however, convenient, to add and subtract $\hat{J}_{8}^2$;
then
the corresponding eigenvalues of Eq.~(\ref{eq:Hcoll})
in the representation $\mathcal{R}=\left(p,\, q\right)$ read:
\begin{align}
\mathcal{E}_{\left(p,\, q\right)}^{\mathrm{rot}} 
= &  M_{\mathrm{sol}} 
    + \frac{J(J+1)}{2I_{1}}  
\cr
  & +\;\frac{C_2(p,q)-J(J+1) -3/4\, Y^{\prime\, 2}}{2I_{2}}
\label{eq:HE}
\end{align}
where $C_2$ denotes SU(3) Casimir and $J$ stands for spin.
The baryon collective eigenfunctions are expressed in terms of
the SU(3) Wigner $D$ functions (see~\cite{Blotz:1992pw,Yang:2010fm}
for details). The right hypercharge imposes a constraint on the
quantization of the chiral soliton,
which for baryons takes the following form: $Y^{\prime}=N_c/ 3$. 
This constraint selects a tower of allowed
rotational excitations of the SU(3) \textit{hedgehog}, which are
identical as in the quark model. This has been considered as a success  
of the collective quantization resulting in a duality between the
chiral soliton picture and the constituent quark model.  
In the case of heavy baryons, as already mentioned, 
$Y^{\prime}=(N_c-1)/ 3$. Then, the lowest rotational excitations
appear to be $(p,q)=(0,\,1)$ (or $\bar{\bm 3}$) with
$S_{\mathrm{L}} = J = 0$ and $(2,\,0)$ (or $\bm 6$) with
$S_{\mathrm{L}} =1$.  

\vspace{0.2cm}

\textit{Explicit SU(3) symmetry breaking}.--The mass splittings in a
heavy baryon multiplet arise from the explicit flavor SU(3) symmetry
breaking caused by the strange current quark mass $m_s$. The
collective Hamiltonian of explicit $\mathrm{SU(3)}_{\mathrm{f}}$
symmetry breaking in the light sector~\cite{Blotz:1992pw} reads:
\begin{align}
H_{\mathrm{br}}  &=  
\alpha \, D_{88}^{(8)} + \beta\,\hat{Y} 
+ \frac{\gamma}{\sqrt{3}}\sum_{i=1}^3D_{8i}^{(8)}\,\hat{J}_{i},
\label{eq:Hsb}
\end{align}
where $\alpha$, $\beta$, and $\gamma$ are given in terms of the
moments of inertia $I_{1,2}$ and $K_{1,2}$, and the pion-nucleon sigma 
term $\Sigma_{\pi N}=(m_u+m_d)\langle
N|\bar{u}u+\bar{d}d|N\rangle/2=(m_u+m_d) \sigma$:  
\begin{align}
\alpha  &=  
- \frac{2 m_s}{3} \sigma  - \beta Y^{\prime}, \;\;\;\;\beta   = -\frac{
         m_s K_{2}}{I_{2}},\cr
\gamma & = \frac{2 m_s K_{1}}{I_{1}} + 2\beta
\label{eq:abr}
\end{align}
In Eq.~(\ref{eq:abr}) we have explicitly included $Y'$, which
is equal to 1 for the light baryons. 

In Ref.~\cite{Yang:2010fm} the dynamical parameters 
$\alpha$, $\beta$ and $\gamma$ have been determined separately 
by using the experimental data for the baryon octet masses, 
the $\Omega$ mass, and the mass of the putative pentaquark
$\Theta^+$, taking into account isospin symmetry breaking 
including the electromagnetic interactions~\cite{Yang:2010id}. The
values of $\alpha$, $\beta$, and $\gamma$ that have been obtained by 
$\chi^2$ minimization \cite{Yang:2010fm} read as follows:  
\begin{align}
\alpha  &=  -255.03\pm5.82 \;{\rm MeV},\cr
\beta & =  -140.04\pm3.20  \;{\rm MeV},
\cr
\gamma  &=  -101.08\pm2.33 \;{\rm MeV},
\label{eq:abrNumber}
\end{align}

When we apply the mean-field approach to heavy baryons, 
$Y'$ is equal to  $Y'=(N_c-1)/3$. 
Analogously, as explained previously, the expressions for the moments
of inertia and $\Sigma_{\pi N}$ need to be modified by a
multiplicative factor of  $(N_c-1)/N_c$. 
Thus, the $m_s$ mass splittings of the heavy baryons 
should be calculated in terms of  $\beta$ and $\gamma$ from 
Eq.~(\ref{eq:abrNumber}), while the value of $\alpha$ should be
modified:
\begin{equation}
\alpha \rightarrow \bar{\alpha}=\frac{N_c-1}{N_c}\alpha.
\label{eq:almod}
\end{equation}

The masses of the anti-triplet and the sextet baryons (without
spin-spin interactions) are then expressed  as:
\begin{align}
M_{B, \mathcal{R}}^Q  
= 
M_{\mathcal{R}}^Q + \delta_{\mathcal{R}}\,Y
\label{eq:M3barM6mass}
\end{align}
where $M_{\mathcal{R}}^Q=m_{Q}  +
\mathcal{E}_{(p,q)}^{\mathrm{rot}}$ is called the center mass
of a heavy baryon in representation $\mathcal{R}$. 
The explicit expressions for $M_{\overline{\bm{3}}}^Q$ and
$M_{\bm{6}}^Q$ are written respectively as: 
\begin{align}
M_{\overline{\bm{3}}}^Q 
&= m_Q + M_{\mathrm{sol}} 
+  \left(\frac{N_c}{N_c-1}\right)\frac{1}{2I_2},
\cr 
M_{\bm{6}}^Q 
&= M_{\overline{\bm{3}}}^Q
+ \left(\frac{N_c}{N_c-1}\right) \frac1{I_1}, 
  \label{eq:mqrep}  
\end{align}
where we have modified the moments of inertia $I_1$ and $I_2$ as
explained above. The term proportional to the hypercharge $Y$ comes
from the explicit $\mathrm{SU(3)}_{\mathrm{f}}$ symmetry breaking in
Eq.~(\ref{eq:Hsb}).  Parameters $\delta_{\overline{\bm 3}}$ and
$\delta_{\bm 6}$ are defined as:
\begin{align}
\delta_{\overline{\bm{3}}}
=
\frac{3}{8}\bar{\alpha}+\beta, \;\;\;\;
\delta_{\bm{6}} 
= \frac{3}{20}\bar{\alpha}+\beta-\frac{3}{10}\gamma. 
\label{eq:d3bard6}
\end{align}

In order to remove the degeneracy between sextet spin 1/2 and 3/2 states, 
we introduce the spin-spin interaction Hamiltonian expressed as:
\begin{align}
H_{LQ} = \frac{2}{3}\frac{\kappa}{m_{Q}\,M_{\mathrm{sol}}}\bm{S}_{\mathrm{L}} 
\cdot  \bm{S}_{Q}  
= \frac{2}{3}\frac{\varkappa}{m_{Q}}
\bm{S}_{\mathrm{L}} \cdot \bm{S}_{Q} 
\label{eq:ssinter}
\end{align}
where $\kappa$ denotes the flavor-independent hyperfine coupling. The
operators ${\bm S}_{\mathrm{L}}$ and ${\bm S}_Q$ represent the spin
operators for the soliton and the heavy quark,
respectively. $M_{\mathrm{sol}}$ has been incorporated into an unknown
coefficient $\varkappa$. The Hamiltonian $H_{LQ}$ does not affect the
$\overline{\bm{3}}$ states, since  in this case $S_{\mathrm{L}}=0$. In
${\bm{6}}$ $S_{\mathrm{L}}=1$, and it couples to $\bm{S}_{Q}$
producing two multiplets $S=1/2$ and $S=3/2$. The respective
splittings read as follows:    
\begin{align}
M_{B,{\bm{6}}_{1/2}}^{Q}
 &=  
M_{B,\bm{6}}^Q\;-\;\frac{2}{3}\frac{\varkappa}{m_{Q}}, 
\cr
M_{B,{\bm{6}}_{3/2}}^{Q}
& = 
M_{B,\bm{6}}^Q\;+\;\frac{1}{3} \frac{\varkappa}{m_{Q}}, 
\label{eq:Csextet}
\end{align}
giving the $3/2-1/2$ splitting  
\begin{align}
M_{B,{\bm{6}}_{3/2}}^{Q}\;-\; M_{B,{\bm{6}}_{1/2}}^{Q}  =
  \frac{\varkappa}{m_{Q}}  .
\label{eq:DCsextet}
\end{align}
That is, $\varkappa$ can be determined by using the center values of
the sextet masses. We list the expressions for the heavy baryon masses
in Table~\ref{tab:1}.  
\begin{table}
\centering{}%
\begin{tabular}{c|ccr|l}
\hline \hline
$\mathbf{\mathcal{R}}_{J}$ 
& $B_{Q}$ 
& $T$ 
& $Y$ 
& $M_{B_{Q}}$
\tabularnewline[0.1em]
\hline 
\multirow{2}{*}{$\mathbf{\overline{3}}_{1/2}$} 
& $\Lambda_{Q}$ 
& $0$ 
& $\frac23$ 
& $ \phantom{-} \frac23\delta_{\overline{\bm{3}}}
+M_{\mathbf{\overline{3}}}^{Q}$
\tabularnewline
& $\Xi_{Q}$ 
& $\frac{1}{2}$ 
& $-\frac{1}{3}$ 
& $-\frac{1}{3}\delta_{\overline{\bm{3}}}
+M_{\mathbf{\overline{3}}}^{Q}$
\tabularnewline
\hline 
\multirow{3}{*}{$\mathbf{6}_{1/2}$} 
& $\Sigma_{Q}$ 
& $1$ 
& ${ \frac{2}{3}}$ 
& $\phantom{-} \frac{2}{3}\delta_{\bm{6}}
-2\varkappa/3m_Q  + M_{\mathbf{6}}^{Q}$
\tabularnewline
& $\Xi_{Q}^{\prime}$ 
& ${ \frac{1}{2}}$ 
& ${ -\frac{1}{3}}$ 
& ${  -\frac{1}{3}\delta_{\bm{6}}
-2\varkappa/3m_Q 
 + M_{\mathbf{6}}^{Q}}$
\tabularnewline
 & $\Omega_{Q}$ 
& $0$ 
& $  -\frac{4}{3}$ 
& $ -\frac{4}{3}\delta_{\bm{6}} -2\varkappa/3m_Q 
 + M_{\mathbf{6}}^{Q}$
\tabularnewline
\hline 
\multirow{3}{*}{$\mathbf{6}_{3/2}$} 
& $\Sigma_{Q}^{\ast}$ 
&  $1$ 
& $ \frac{2}{3}$ 
& $ \phantom{-}\frac{2}{3}\delta_{\bm{6}}
+\varkappa/3m_Q  + M_{\mathbf{6}}^{Q}$
\tabularnewline
 & $\Xi_{Q}^{\ast}$ 
& ${ \frac{1}{2}}$ 
& ${ -\frac{1}{3}}$ 
& $ -\frac{1}{3}\delta_{\bm{6}}
+\varkappa/3m_Q +M_{\mathbf{6}}^{Q}$
\tabularnewline
 & $\Omega_{Q}^{\ast}$ 
& $0$ 
& ${ -\frac{4}{3}}$ 
& $-\frac{4}{3}\delta_{\bm{6}}
+\varkappa/3m_Q 
+M_{\mathbf{6}}^{Q}$
\tabularnewline
\hline \hline
\end{tabular}
\caption{Expressions for the masses of the heavy baryons 
\label{tab:1}}
\end{table}

\textit{Model-independent relations}.--The mass formulae given in  
Table~\ref{tab:1} imply relations that do not depend upon actual
values of the model parameters -- so called \emph{model-independent
  relations}.\footnote{The term \textit{model-independent relations} in
  the present context has been first used in
  Ref.~\cite{Adkins:1984cf}. It refers to the fact that the operators
  that appear \textit{e.g} in (\ref{eq:Hsb}) follow from the \textit{
    hedgehog} symmetry, rather than from a specific model.} 
   An immediate consequence of the mass formulae of
Table~\ref{tab:1} is the equal mass splittings separately in the
$\overline{\bm{3}}$ and $\bm{6}$. Note that the mass splittings are
independent of the spin and of the heavy quark mass. These relations
are indeed very well satisfied.\footnote{All model independent relations are checked 
using the data from \cite{Agashe:2014kda} quoted in Tables~\ref{tab:2} and \ref{tab:3}.
For isospin multiplets an average mass is used.}
For the $\bar{\bf 3}$ we have (in
MeV): 
\begin{align}
-\delta_{\overline{\bm 3}}
=\left. 182.9 \pm 0.3
\right|_{\Xi_c-\Lambda_c}
=\left. 173.6  \pm 0.7
\right|_{\Xi_b-\Lambda_b}, 
\label{eq:deltabar3}
\end{align}
which is satisfied with 7~\% accuracy. In the case of  the $\bm{6}$ we 
have more relations (in MeV): 
\begin{align}
 -\delta_{\bm{6}}& 
= \left. 123.3 \pm 2.1
  \right|_{\Xi^{\prime}_c-\Sigma_c}
= \left. 118.4 \pm 2.7
  \right|_{\Omega_c - \Xi^{\prime}_c} 
\cr
& 
= \left. 127.8 \pm 0.8
  \right|_{\Xi^{\ast}_c-\Sigma^{\ast}_c}
= \left. 120.0 \pm 2.0
  \right|_{\Omega^{\ast}_c - \Xi^{\ast}_c}
\cr
&
= \left. 121.6 \pm 1.3
  \right|_{\Xi^{\prime}_b-\Sigma_b}
= \left. 113.0 \pm 1.9
  \right|_{\Omega_b - \Xi^{\prime}_b}
\cr 
&
= \left. 121.7 \pm 1.3
  \right|_{\Xi^{\ast}_b-\Sigma^{\ast}_b}.
 \label{eq:delta6}
\end{align}
We see that the equality of splittings is quite accurate (at the 6~\%
level). From the spread of splittings in Eq.~(\ref{eq:delta6}), 
we can make the first prediction of the mass of $\Omega^{\ast}_b$ that
is not yet measured, taking as an input the experimental mass of
$\Xi^{\ast}_b$:  
\begin{equation}
M_{\Omega_{b}^{\ast}}= (6068 - 6083)~\mathrm{MeV}.
\label{eq:Omstb1}
\end{equation}

Using the mass formulae presented in Table~\ref{tab:1}, we are able to
determine the centers of the heavy baryon masses:
\begin{align}
M_{\overline{\bm{3}}}^Q
 =  
\frac{M_{\Lambda_{Q}}+2M_{\Xi_{Q}}}{3},\;\;
M_{\bm{6}}^{Q} 
=
\frac{M_{\bm{6}_{1/2}}^{Q} + 2M_{\bm{6}_{3/2}}^{Q}}{3}
\label{eq:M3barM6}
\end{align}
where
\begin{align}
M_{\bm{6}_{1/2}}^{Q} 
 & = \frac{3M_{\Sigma_{Q}} + 2M_{\Xi_{Q}^{\prime}}
   + M_{\Omega_{Q}}}{6},
\cr 
M_{\bm{6}_{3/2}}^{Q} 
 & = \frac{3M_{\Sigma_{Q}^{\ast}}
    + 2M_{\Xi_{Q}^{\ast}}
    + M_{\Omega_{Q}^{\ast}}}{6}, 
\label{eq:M61M63}
\end{align}
with $M_{\bm{6}_{1/2,\,3/2}}^{Q}$ given in Eq.~(\ref{eq:Csextet}).
Equation~(\ref{eq:M61M63}) can not be used in the $b$ sector, because 
we do not know $M_{\Omega_{b}^{\ast}}$. Fortunately, we can determine
the centers of the multiplets without invoking  $\Omega_Q$ masses. Defining
\begin{align}
S(\Sigma_Q)
&= \frac{M_{\Sigma_{Q}}+ 2 M_{\Sigma_{Q}^{\ast}}}{3}
 = M_{\bm{6}}^{Q} +\frac{2}{3}\delta_{\bm{6}},
\cr 
S(\Xi_Q)
&= \frac{M_{\Xi^{\prime}_{Q}}+ 2 M_{\Xi_{Q}^{\ast}}}{3}
 = M_{\bm{6}}^{Q} -\frac{1}{3}\delta_{\bm{6}}.    
\end{align}
we have
\begin{align}
M_{\bm{6}}^{Q}=\frac{S(\Sigma_Q)+2S(\Xi_Q)}{3}.
\label{eq:M6QS}
\end{align}
For the sextet in the $c$ sector, Eqs.~(\ref{eq:M3barM6},
\ref{eq:M6QS}) can be regarded as model-independent relation:   
\begin{align}
M_{\bm{6}}^{c} 
= \left. 2579.6 \pm 0.4 \right|_{\rm Eq.(\ref{eq:M3barM6})} 
= \left. 2580.8 \pm 0.5 \right|_{\rm Eq.(\ref{eq:M6QS})} 
\label{eq:relM6c}
\end{align}
in MeV. Relation~(\ref{eq:relM6c}) is fulfilled with
unprecedented accuracy. For the ${\bm{3}}$ and for the {\bf 6}
in the $b$ sector we have: 
\begin{align}
M_{\overline{\bm{3}}}^c 
& = \left. (24087.4\pm0.2)
    \right|_{\mathrm{Eq.(\ref{eq:M3barM6})}}\; \mathrm{MeV}, 
\cr  
M_{\overline{\bm{3}}}^b 
& = \left. (5735.2\pm0.4)
    \right|_{\mathrm{Eq.(\ref{eq:M3barM6})}}\;\mathrm{MeV}, 
\cr  
M_{\bm{6}}^{b}
& = \left. (5908.0\pm0.3)
    \right|_{\mathrm{Eq.(\ref{eq:M6QS})}}\;\mathrm{MeV}.
\label{eq:Mcenter}
\end{align}

Apart from equal splittings in {\bf 6} (\ref{eq:delta6}), mass formulae
of Tab.~\ref{tab:1} admit a sum rule:
\begin{align}
M_{\Omega_{Q}^{\ast}} 
= 
 2M_{\Xi^{\prime}_{Q}}
 +M_{\Sigma_{Q}^{\ast}} 
 -2 M_{\Sigma_{Q}}.
\label{eq:OMpred}
\end{align}
Equation~(\ref{eq:OMpred}) yields $(2764.5\pm3.1)$~MeV for
$M_{\Omega_{c}^{\ast}}$, which is 1.4~MeV below the experiment, 
and predicts 
\begin{align}
M_{\Omega_{b}^{\ast}}=(6076.8\pm2.25)\;\mathrm{MeV},
\label{eq:Omstb2}
 \end{align}
which falls in the range of Eq.~(\ref{eq:Omstb1}). 
  
Equation~(\ref{eq:mqrep}) provides yet another model independent
relation, allowing one to determine the moment of inertia $I_1$
either from the $c$ or from the $b$ sector: 
\begin{align}
\frac{1}{I_1}=\frac{2}{3}(M_{\mathbf{6}}^Q -
  M_{\mathbf{\overline{3}}}^Q)  
=\left. 114.7\right|_c=\left. 115.2\right|_b
\label{eq:I1Q}
\end{align}
in MeV.~\footnote{A similar relation is found in
  Ref.~\cite{Momen:1993ax} with a different factor and 
  in a different 
  context.}   
The reason for the equality of splittings between the multiplet
centers can be traced back to the fact that it comes only from the
energy of the rotational excitations, which are flavor-blind in the
present approach. Moreover, the effects of the 
$\mathrm{SU(3)}_{\mathrm{f}}$ symmetry breaking are simply the same
both for the charm and bottom baryons, since they are solely due to
the presence of the light quarks inside a heavy baryon. The relation
of Eq.~(\ref{eq:I1Q}) is indeed very accurate, but it undershoots by
30~\% the value of $1/I_1$ extracted from the light sector equal to
160~MeV.  

Another set of model-independent relations is not directly related to
the specifics of the soliton model, but provides a test of our
assumption concerning the spin interactions of
Eq.~(\ref{eq:DCsextet}):
\begin{align}
\frac{\varkappa}{m_c}
&= \left. 64.5\pm0.8 \right|_{\Sigma_c}
 = \left. 69.1\pm2.1 \right|_{\Xi_c} 
 = \left. 70.7\pm2.6 \right|_{\Omega_c} 
\cr
\frac{\varkappa}{m_b}
&= \left. 20.2\pm1.9 \right|_{\Sigma_b}
 = \left. 20.3\pm0.1 \right|_{\Xi_b}.
\label{eq:spintest}
\end{align}
(in MeV). Equation~(\ref{eq:spintest}) provides yet another prediction for the
$\Omega^{\ast}_b$ mass 
\begin{align}
M_{\Omega_{b}^{\ast}}
= M_{\Omega_{b}}+\frac{\varkappa}{m_b}
= (6068.3\pm2.1)~\mathrm{MeV}
\label{eq:Omstb3}
\end{align}
in good agreement with Eq.~(\ref{eq:Omstb2}).
From the ratios of the spin splittings (\ref{eq:spintest}) we can determine
the ratio of the heavy-quark masses 
\begin{align}
\frac{m_c}{m_b}=0.29 - 0.31.
\label{eq:cbratio}
\end{align}
The experimental values of the $\overline{\rm MS}$ heavy quark masses  
lead to $m_c/m_b=0.305$ inserted, where both masses $m_Q$ are
evaluated at the renormalization point
$\mu=m_Q$~\cite{Agashe:2014kda}.  
Of course heavy quark masses in the effective models, like the one
considered in this paper, may differ from the QCD masses. It is
therefore encouraging that we get the mass ratio close to the ratio of
the QCD masses. 
\vspace{0.2cm}

\textit{Masses of heavy baryons}.--Having determined $\varkappa/m_Q$,
using the numerical values of $\bar{\alpha}$, $\beta$ and
$\gamma$ from Eqs.~(\ref{eq:abrNumber}) and (\ref{eq:almod}), 
we can predict the masses  
of the lowest-lying ($\overline{\bm 3}$ and $\bm 6$) heavy baryons.   
As we have already mentioned the determination of $I_1$ from the heavy
quark sector and from the light sector differ by 30~\%. Therefore in
the following we shall use the center masses $M_{\overline{\bm 3}}^Q$
and $M_{\bm 6}^Q$ as given by Eqs.~(\ref{eq:relM6c},\ref{eq:Mcenter}).  

As a first check let us compare the values of $\delta$ parameters
determined from the  light sector through Eqs.~(\ref{eq:d3bard6}): 
\begin{align}
\delta_{\overline{\bm{3}}}& = -203.8\pm3.5\;{\rm MeV}, \notag \\
  \delta_{\bm{6}}& = -135.2\pm3.3\;{\rm MeV},
\label{eq:deltamodel}
\end{align}
with the values  following from the heavy sector
given in Eqs.~(\ref{eq:deltabar3},\ref{eq:delta6}). We see
that the light sector values (\ref{eq:deltamodel})
underestimate the heavy quark
determination (\ref{eq:deltabar3}) and (\ref{eq:delta6}) by approximately
13~\%. 
Interestingly, the ratio $\delta_{\overline{\bm
    3}}/\delta_{\bm{6}}=1.5$ is almost exactly equal to the ratio of 
the average splittings as given in Eqs.~(\ref{eq:deltabar3},\ref{eq:delta6}). 
Accuracy of the predictions given in
Eq.~(\ref{eq:deltamodel}) deserves a comment.  
Equal splittings in $\overline{\bm 3}$
or $\bm 6$ are analogous to the Gell-Mann--Okubo mass formulae
for the  light baryon decuplet and follow solely from the SU(3)$_{\rm f}$ group
properties. However, the relation between the splittings in
$\overline{\bm 3}$ and $\bm 6$ is a complicated dynamical
question. The fact that chiral dynamics with an input from the
light-baryon sector alone, reproduces $\delta_{\overline{\bm 3}}$ and
$\delta_{\bm{6}}$ with good accuracy is therefore by far not trivial. 
The fact that the ratio $\delta_{\overline{\bm{3}}}/\delta_{\bm{6}}$
is even better reproduced suggests a multiplicative modification of
these parameters by a common factor, which may be due \textit{e.g.} to
a slight change of $m_s$ in the heavy baryon environment. 

In the following we shall use $M_{\bm 6}^{c}=2580.8$~MeV, and 
 Eq.~(\ref{eq:Mcenter}) for the multiplet 
centers, $\delta$ parameters from Eq.~(\ref{eq:deltamodel}),
and the average values for the hyperfine splittings: 
$\varkappa/m_c=(68.1\pm1.1)$~MeV and $\varkappa/m_b=(20.3\pm1.0)$~MeV.  
In order to quantify the quality of the predictions we introduce
deviation  $\xi_Q =  (M_{\mathrm{th}}^{B_Q} -
  M_{\mathrm{exp}}^{B_Q})/M_{\mathrm{exp}}^{B_Q}$,     
where $M_{\mathrm{th}}^{B_Q}$ represents the prediction of the present
work, whereas $M_{\mathrm{exp}}^{B_Q}$ stands
for the experimental value. The
results presented in Tables ~\ref{tab:2} and \ref{tab:3} are
in remarkable agreement with the experimental data within $0.7\,\%$ or
less. Note that the uncertainties in Tables ~\ref{tab:2} and
\ref{tab:3} include those from the multiplet centers, $\varkappa/m_Q$,
$\alpha$, $\beta$, and $\gamma$.
\begin{table}
\begin{centering}
\begin{tabular}{c|cccr}
\hline \hline
$\mathbf{\mathcal{R}}_{J}^{Q}$ 
& $B_{c}$ 
& Mass 
& Experiment~\cite{Agashe:2014kda}
& Deviation $\xi_c$
\tabularnewline[0.1em]
\hline 
\multirow{2}{*}{$\mathbf{\overline{3}}_{1/2}^{c}$} 
& $\Lambda_{c}$ 
& $2272.5 \pm 2.3$
& $2286.5 \pm 0.1$
& $-0.006$
\tabularnewline
& $\Xi_{c}$ 
& $2476.3 \pm 1.2$
& $2469.4 \pm 0.3$
& $0.003$
\tabularnewline
\hline 
\multirow{3}{*}{$\mathbf{6}_{1/2}^{c}$} 
& $\Sigma_{c}$ 
& $2445.3 \pm 2.5$
& $2453.5 \pm 0.1$
& $-0.003$
\tabularnewline
& $\Xi_{c}^{\prime}$ 
& $2580.5 \pm 1.6$
& $2576.8 \pm 2.1$
& $0.001$
\tabularnewline
& $\Omega_{c}$ 
& $2715.7 \pm 4.5$
& $2695.2 \pm 1.7$
& $0.008$
\tabularnewline
\hline 
\multirow{3}{*}{$\mathbf{6}_{3/2}^{c}$} 
& $\Sigma_{c}^{\ast}$ 
& $2513.4 \pm 2.3$
& $2518.1 \pm 0.8$
& $-0.002$
\tabularnewline
& $\Xi_{c}^{\ast}$ 
& $2648.6 \pm 1.3$
& $2645.9 \pm 0.4$
& $0.001$
\tabularnewline
& $\Omega_{c}^{\ast}$ 
& $2783.8 \pm 4.5$
& $2765.9 \pm 2.0$
& $0.006$
\tabularnewline
\hline \hline
\end{tabular}
\par\end{centering}
\caption{The results of the masses of the charmed baryons in
  comparison with the experimental data.}
\label{tab:2}
\end{table}

\begin{table}[htp]
\centering{}\vspace{3em}%
\begin{tabular}{c|cccr}
\hline\hline 
$\mathbf{\mathcal{R}}_{J}^{Q}$ 
& $B_{b}$ 
& Mass
& Experiment~\cite{Agashe:2014kda}
& Deviation $\xi_b$
\\
\hline 
\multirow{2}{*}{$\mathbf{\overline{3}}_{1/2}^{b}$} 
& \textcolor{black}{$\Lambda_{b}$} 
& $5599.3 \pm 2.4 $
& $5619.5 \pm 0.2$ 
& $-0.004$
 \tabularnewline
& \textcolor{black}{$\Xi_{b}$} 
& $5803.1 \pm 1.2 $
& $5793.1 \pm 0.7 $  
& $0.002$
 \tabularnewline
\hline 
\multirow{3}{*}{$\mathbf{6}_{1/2}^{b}$} 
& \textcolor{black}{$\Sigma_{b}$} 
& $5804.3 \pm 2.4 $
& $5813.4 \pm 1.3$  
& $ -0.002$
\tabularnewline
& \textcolor{black}{$\Xi_{b}^{\prime}$} 
& $5939.5 \pm 1.5 $
& $5935.0 \pm 0.05$ 
& $ 0.001$
\tabularnewline
& \textcolor{black}{$\Omega_{b}$} 
& $6074.7 \pm 4.5 $
& $6048.0 \pm 1.9$  
& $ 0.004$
 \tabularnewline
\hline 
\multirow{3}{*}{$\mathbf{6}_{3/2}^{b}$} 
& \textcolor{black}{$\Sigma_{b}^{\ast}$} 
& $5824.6 \pm 2.3 $
& $5833.6 \pm 1.3$  
& $-0.002$
\tabularnewline
&\textcolor{black}{$\Xi_{b}^{\ast}$} 
& $5959.8 \pm 1.2 $
& $5955.3 \pm 0.1$ 
& $ 0.001$
 \tabularnewline
& \textcolor{black}{$\Omega_{b}^{\ast}$} 
& $6095.0 \pm 4.4 $
& $-$
& $-$
\tabularnewline
\hline \hline
\end{tabular}
\caption{The results of the masses of the bottom baryons in comparison
with the experimental data.}
\label{tab:3}
\end{table}

In the last row of Table~\ref{tab:3}, the mass of the
$\Omega^{\ast}_b$ is predicted:
\begin{align}
M_{\Omega^{\ast}_b} = (6095.0 \pm 4.4\pm 24)\; \mathrm{MeV},
\label{eq:Omstb4}
\end{align}
where $\pm 24$~MeV corresponds to the overall accuracy of 
of the model which we assume to be, as for the other
bottom states, within the 0.4\% range.
This result lies slightly above the other predictions obtained in
Eqs.~(\ref{eq:Omstb1}, \ref{eq:Omstb2}, \ref{eq:Omstb3}). 

\textit{Summary and Outlook.}--In the present Letter, we have applied
 a pion mean-field approach with \textit{hedgehog} symmetry to the
description of the heavy baryon masses, which essentially assumes
that the heavy quark ($c$ or $b$) is surrounded by a pion mean field
or a light-quark soliton produced from the $N_c-1$ \textit{valence} 
quarks. This assumption leads to a number of model-independent
predictions: (i) the soliton quantization forces heavy baryons to have 
the following SU(3)$_{\mathrm{f}}$ structure: $\overline{\bm{3}}$
 with spin 1/2, and the $\bm{6}$ with spin 1/2 and 3/2   
with approximate degeneracy of the sextets, (ii)  equal mass
splittings within the multiplets given in
Eqs.~(\ref{eq:deltabar3},\ref{eq:delta6}) that do not depend on the  
heavy quark mass, (iii)  equal splittings between
$\overline{\bm{3}}$ and $\bm{6}$ for the $c$ and $b$ sectors
(\ref{eq:I1Q}), and (iv) the sum rule that allows to calculate the
mass of $\Omega^{\ast}_Q$ in Eq.~(\ref{eq:OMpred}). 

We have completed the model by adding  the hyperfine interaction that
is inversely proportional to the heavy quark mass in
Eq.~(\ref{eq:ssinter}). This assumption proved to very accurate 
as shown in Eq.~(\ref{eq:cbratio}) and the pertinent coefficient has
been determined from the mass splittings in Eq.~(\ref{eq:spintest}).

Next, we have used the three parameters extracted from the light
baryon sector in Eq.~(\ref{eq:abrNumber}) to calculate heavy baryon
masses. The soliton has been assumed to be exactly the same as in the 
case of the light baryons with one exception:  all \textit{moments of
  inertia} have been rescaled by a factor of $(N_c-1)/N_c$, because
 there are $N_c-1$ rather than $N_c$ \textit{valence quarks} in a heavy baryon. 
 With this modification, the predictions for the heavy
baryon masses turned out to be within 0.5~\% range when 
compared with the experiment data. Unfortunately, the splitting of the
multiplet centers of mass equal to $3/2I_1$ turned out to be 30~\% off
the light baryon prediction.  
This result suggests that, apart from the $(N_c-1)/N_c$ factor,
moments of inertia be further modified in the presence of the heavy
quark. These modifications to the large extent cancel in the ratios
that enter Eq.~(\ref{eq:abr}) except for the $\Sigma_{\pi N}$ term  
in the definition of $\alpha$. We have checked, however, that varying
$\Sigma_{\pi N}$ by $\pm 30$~\% changes $\delta_{\overline{\bm{3}}}$ by 
$\pm 8.6$~MeV, and $\delta_{\bm 6}$ by $\pm3.5$~MeV. Such a change will
not affect the quality of the predictions for the heavy baryon masses
presented in this Letter.  

Finally, we have presented four different predictions of the
$\Omega^{\ast}_b$ mass: (i) from the spread of the $m_s$ splittings in
${\bm 6}$ (\ref{eq:Omstb1}), (ii) from the $\Omega^{\ast}_{Q}$ sum
rule (\ref{eq:Omstb2}), (iii) from the hyperfine splittings
(\ref{eq:Omstb3}), and finally we have given model prediction in
Eq.~(\ref{eq:Omstb4}). All these estimates are consistent and point to
the value of $M_{\Omega^{\ast}_b}$ within the range of
Eq.~(\ref{eq:Omstb1}). Only the prediction of Eq.~(\ref{eq:Omstb4})
based on the parameters of the light-baryons sector was obtained
to be slightly higher than the other ones. We anticipate that the mass
of $\Omega^{\ast}_b$ will soon be measured at the Large Hadron
Collider (LHC).  

The present work has clear physical implications. The mean fields of
the pion play indeed a crucial role in explaining not only
the masses of the lowest-lying baryons in the light-quark sector but
also those of the heavy baryons. The feedback of the heavy quark on
the light sector may be of order of 30 \%, but it largely cancels
in the heavy-baryon splittings. This aspect of the pion mean field
deserves further study. 

\vspace{0.5cm} 
We are grateful to A. Hosaka, S.H. Kim, and U. Yakhshiev for
valuable discussions. The work of H.-Ch.K. was supported by Basic
Science Research Program through the National Research Foundation of
Korea funded by the Ministry of Education, Science and Technology
(Grant Number: NRF-2015R1D1A1A01060707).

\end{document}